\documentclass[aps,prb,twocolumn,superscriptaddress]{revtex4-1}
\usepackage{amsmath,amssymb}
\usepackage{amsfonts}
\usepackage{bbold}
\usepackage{graphicx}
\usepackage{color}
\usepackage{float}
\usepackage{hyperref}
\usepackage{subcaption}
\usepackage{epstopdf}

\graphicspath{{figs/}}

\begin{document}
\title{Percolation transition in two dimensional electron gas: A quantum cellular automaton model}
\author{M. N. Najafi}
\affiliation{Department of Physics, University of Mohaghegh Ardabili, P.O. Box 179, Ardabil, Iran}
\begin{abstract}
A new type of disorder-driven electronic percolation transition is found for two-dimensional electron gas (2DEG), based on a quantum cellular automaton model. This transition is shown to be accompanied with a metal-insulator transition, as well as a singularity in the electronic compressibility. To this end, the electronic system which is assumed to be in contact with an electronic reservoir, is meshed by using of the phase coherence length $\zeta_{\phi}$ as an extent which divides the spatial dynamics of the electrons into two separate regimes and controls the localization of electrons. For the scales much smaller than $\zeta_{\phi}$ the treatment is quantum mechanical, whereas for the scales much larger than $\zeta_{\phi}$ the picture of semi-classical transport works and the classical Mote Carlo method is used. Thomas-Fermi-Dirac (TFD) theory is employed to find the dependence of the free energy of each cell on the temperature ($T$), the (charged) disorder strength ($\Delta$) and the charge content of the cell, and a cellular automaton model for transporting the electrons between the cells is developed. At the transition line (in the $T-\Delta$ space) the geometrical (e.g. correlation length) quantities also diverge and some power-law behaviors emerge with some critical exponents in agreement with the Gaussian free field (GFF) and the percolation theory. In the percolating side the spanning cluster probability (SCP) (as a realization of the conductivity) has a decreasing behavior in terms of the temperature which is the characteristics of the metallic phase. Our simple model yields the important features of the experimental observations, e.g. the singularity in the conductivity in some critical density and also the universality (non-universality) of the metal-insulator transition (MIT) for the small (large) disorders in 2DEG. A $T-\Delta$ phase diagram of the electron gas is drawn in which along with the mentioned transition line, a zero-heat capacity line is also observed in which the system becomes unstable.
\end{abstract}
\maketitle
\section{Introduction}
\begin{figure*}
	\begin{subfigure}{0.45\textwidth}\includegraphics[width=\textwidth]{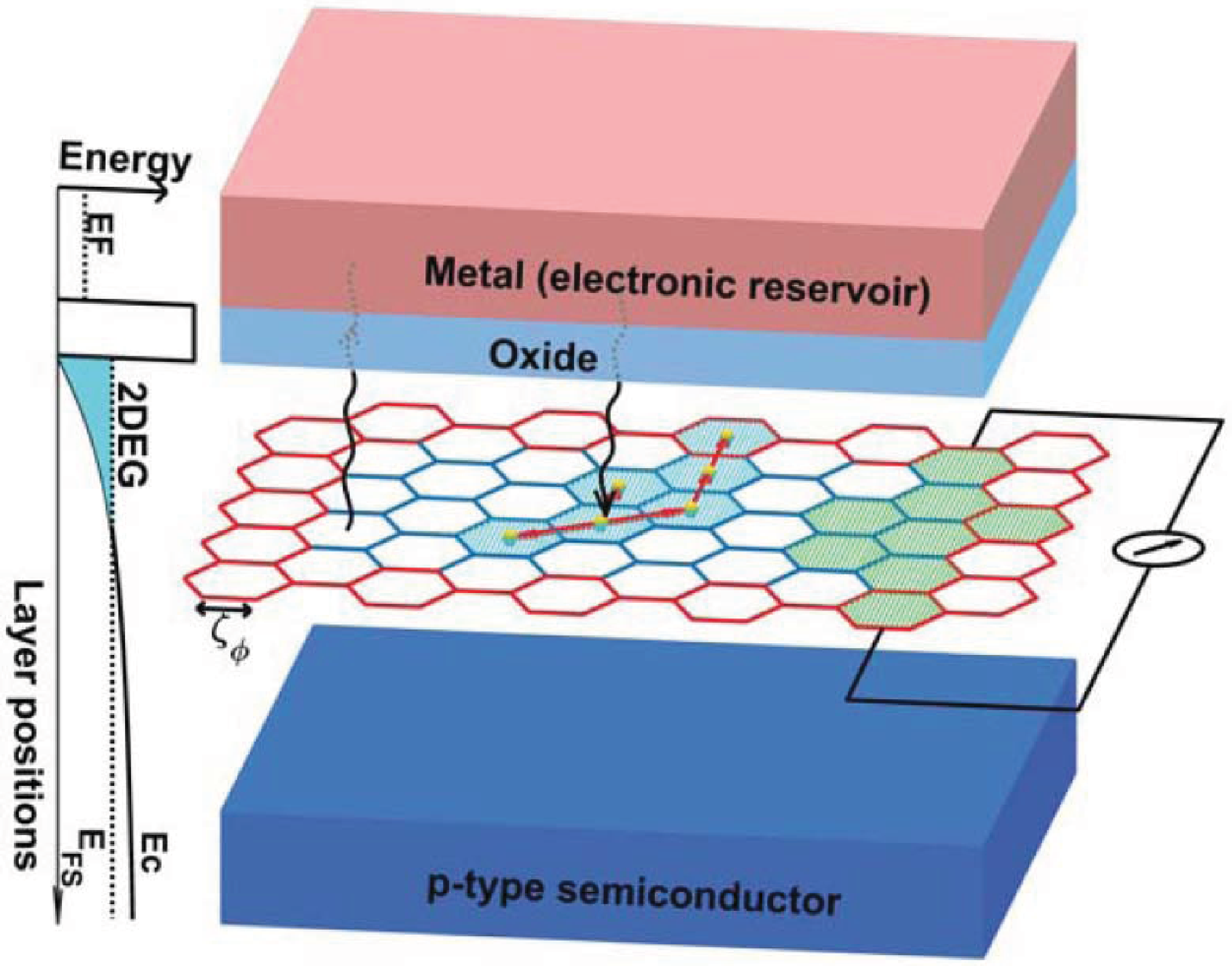}
		\caption{}
		\label{fig:SchematicSetUp}
	\end{subfigure}
	\begin{subfigure}{0.5\textwidth}\includegraphics[width=\textwidth]{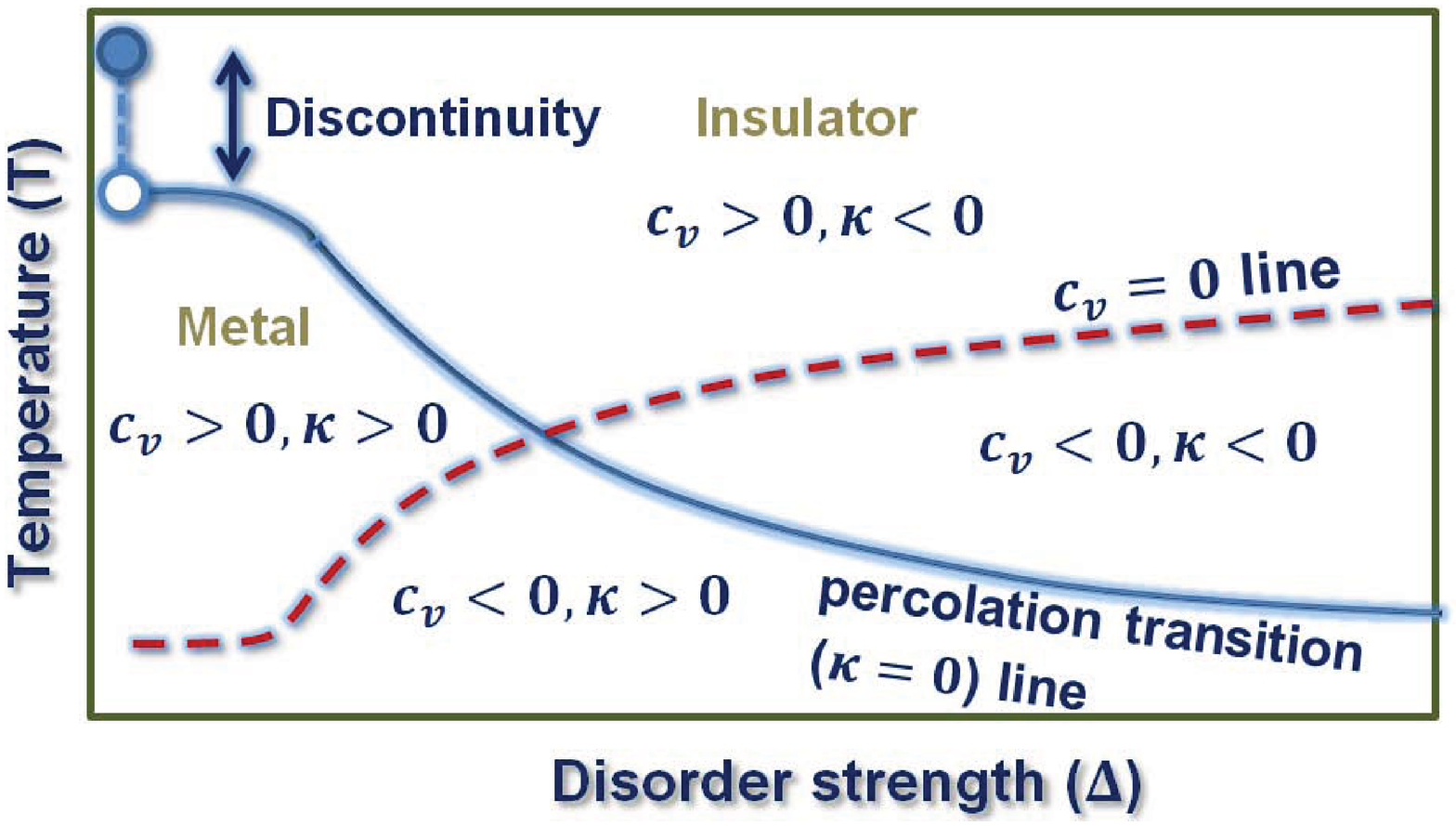}
		\caption{}
		\label{fig:PhaseSpace}
	\end{subfigure}
	\caption{(a) A schematic representation of a MOS system with a 2DEG in the inversion layer. The semiconductor is grounded and the metal is negatively biased with $V<0$ (sufficiently negative to form an inversion layer). The 2DEG has mashed by many hexagons of the linear size $\zeta_{\phi}$. Inside the hexagons we have a pure quantum electron gas and the transfer between the cells occur semi-classically. The metal and the semiconductors play the roles of electronic reservoirs. The electron that is transmitted from the metal to the 2DEG leads to some electronic transmissions in the neighboring cells, resulting to an affected area which is composed of the \textit{toppled} cells (shadowed area). An spanning \textit{toppled} area contributes to the conductivity, as has been shown in the figure. The electrons can be enter and leave the 2DEG at any random point (with energy considerations), and can be dissipated from the boundaries which have been shown in red. (b) The total (schematic) $T-\Delta$ phase diagram with zero-$\kappa$ (compressibility) (blue full line) and zero-$c_v$ (red broken line). The zero-$\kappa$ is along with the geometrical percolation transition. The sign of $\kappa$ and $c_v$ in each phase has been shown. For the $c_v>0, \kappa>0$ phase, we have $\frac{\text{d}\sigma}{\text{d}T}<0$ ($\sigma\equiv$ conductivity) which is a metallic behavior. A discontinuity is also seen from $\Delta=0$ to $\Delta=0^+$ showing that the problem is not preturbative with respect to $\Delta$.}
	\label{fig:firstface}
\end{figure*}
Many classical and quantum systems have the potential to be described in terms of the percolation theory in some limits \cite{SaberiPercolation}. Despite its very simple rules, this theory has successfully been applied to describe a large variety of natural \cite{najafi2015water,SaberiMoon}, social \cite{Newman}, and quantum \cite{Girvin,Sarma} systems. The description of insulating-non-insulating phase transitions in terms of the percolation theory has been done in some systems, like quantum Hall effect systems~\cite{Girvin} and metal-insulator transition (MIT) in two-dimensional electron gas (2DEG)~\cite{Sarma}. In the latter case, in an essentially classical scheme, it is claimed that the disorder-induced \textit{valleys and mountains} created by unscreened coulomb potential at low densities \cite{Sarma} is responsible (or has the dominant role) for the experimentally observed MIT in 2DEG systems~\cite{Abrahams,Kravchenko,Kravchenko1,AbrahamsAnderson}. The percolation approach due to its generality and simpleness sounds very promising in such quantum systems. The existence of such \textit{valleys and mountains} have also been suggested in the electron-hole puddles of graphene to describe the minimum conductivity of graphene~\cite{Rossi}.\\
The percolation theory, when is mixed by the cellular automaton models is proved to be very powerful in describing the natural systems~\cite{Chopard,Noest}. The classical example is the sandpiles on the percolation lattices which has some relations with the propagation of fluid in the reservoirs~\cite{najafi2015water}. The easiest way to generalize this concept to the quantum systems is to use the notion of phase-relaxation length $\zeta_{\phi}$ above which the electronic transport is classical, i.e. the system can be meshed by means of $\zeta_{\phi}$. A cellular automaton model can therefore be designed for the transport of electrons with spatial scales below and above $\zeta_{\phi}$. Consider for example a highly biased metal-oxide-semiconductor (MOS) junction with an inversion layer in which a 2D electron gas is formed whose density is controlled by the bias voltage~\cite{Kravchenko}. In this system the metal and semiconductor have the role of electronic reservoirs from which the electrons can be transferred to the inversion layer, i.e. 2DEG and vice versa. The localization of electrons in such a system is roughly controlled by $\zeta_{\phi}$. In Fig. \ref{fig:SchematicSetUp} this system has schematically been shown, in which the 2DEG has been meshed by some cells (hexagons, to be most rotationally symmetric and should not be confused with a real lattice) of the linear scale $\zeta_{\phi}$. The aim of the present paper is to develop a cellular automaton model with local transition rules based on the local chemical potential. The electrons, when transmitted from the electronic reservoirs to the inversion layer, respect to some simple automaton rules for traveling to the neighboring cells (hexagons), resulting to a chain of in-plane transmissions. By coloring the cells in which a transmission has occurred, a colored area results. A spanning colored area means that some electrons have traveled throughout the 2D sample and contributed to the in-plane conductivity. This method, when compared with the other percolation methods, e.g. two-component effective medium theory~\cite{Sarma}, is proved to be very rich and powerful. It has the ability to bring temperature, disorder and inter-particle interactions simultaneously on an footing in the calculations and shows various phases, reflecting the rich internal structure of the system. Our main observation of the paper, namely that in the vanishing inter-particle interaction, and in the diffusive phase the system experiences a transition from the localized states (in which the electrons cannot travel throughout the sample) to the extended states (metallic states in the sense that the conductivity is a decreasing function of the temperature), shows that the system can have a MIT which is not interaction-driven (like Wigner-Mott systems \cite{CamjayiKotliar,Radonjic,Vucicevic,Byczuk}), nor Anderson-localization type \cite{GoldLocalization}, instead a \textit{finite temperature disorder-driven percolation transition} (for 2D MIT see~\cite{2DMetals}). We name this behavior as \textit{semi-classical localization} of electrons. The Thomas-Fermi-Dirac theory has been used as the quantum model for the electron gas inside the cells (hexagons) and the inter-cell transition probabilities have been considered to be according to the classical Boltzmann weights, for which the Metropolis Monte-Carlo technique has been employed. For any time, the status of the electrons is updated according to the state of the electrons just in the previous time step. Our simple model yields the main features of the experimental observations, like the singularity in the conductivity in some disorder-dependent critical density $n_c$ and also the universality (non-universality) of the metal-insulator transition (MIT) for the small (large) disorders in 2DEG.\\
The Fig.~\ref{fig:PhaseSpace} has been shown to summarize our main results, in which the phase digram of the system has been schematically drawn in terms of temperature $T$ and disorder strength $\Delta$ for zero inter-particle interaction. The full (blue) line shows the second order percolation transition above which the system is localized and under which the system is extended. At this line the compressibility of the system diverges, signaling an instability of the system. In the extended phase, the conductivity has a decreasing behavior in terms of temperature which is the characteristics of the \textit{metallic phase}. The electron density as well as the chemical potential show also some singular behaviors at this line. The dashed (red) line shows the zero-heat capacity ($C_v=0$) line below which $C_v<0$ in which the electron gas is unstable. The vertical full line shows that the system shows a discontinuity from $\Delta=0$ to positive $\Delta$ values. These lines meet each other at some critical value $\Delta^*$. \\
The paper has been organized as follows. In SEC~\ref{statement} we state the problem and its motivation. The free energy of the coherent electron gas inside the cells is calculated in this section. Local and global properties of the model are investigated in SECs~\ref{local} and \ref{global} respectively.

\section{The statement of the problem, Energy functional and the electronic transition rules}\label{statement}
When a quantum system is in the diffusive regime, the dynamics of electrons is divided to two scales relative to the phase relaxation length $\zeta_{\phi}\equiv \sqrt{D\tau_{\phi}}$ in which $D$ is the diffusion coefficient and $\tau_{\phi}$ is the phase relaxation time associated with inelastic or spin-flip scattering up to which the electrons retain their coherence, i.e. the quantum phase is maintained up to $t=\tau_{\phi}$. Therefore the spatial dynamics of the electrons is divided to two spatial ($r$) scales : $l \ll r \ll \zeta_{\phi}$ and $r\gg \zeta_{\phi}$ in which $l$ is the mean free path due to the electron-electron or the electron-phonon interactions. For $r\gg \zeta_{\phi}$ the electron dynamics is semi-classical, since quantum fluctuations become weak and one can use classical Boltzmann transport equation \cite{altshuler}. The existence of such a spatial scale has proved to be useful in many situations and physical interpretation of some phenomenon, e. g. the self averaging of quantum systems. An example is the measured conductance of a disordered GaAs sample in high and low temperatures \cite{Flensberg}. In this approach one subdivides the system into many cells with the linear sizes $\Delta L\sim\ \zeta_{\phi}$ which is surely temperature dependent, i.e. $\zeta_{\phi}(T)$. The electron gas inside the cells should be treated quantum mechanically, whereas the transport between the cells is semi-classical. The cells should be considered most symmetric, i.e. circles in 2DEG (spheres in 3D) which has been estimated by hexagons in the present work. The meshed space has schematically drawn in Fig. \ref{fig:SchematicSetUp} in contact with some electronic reservoirs (metal and semiconductor) from which the electrons can cross out. The electrons can enter the 2DEG from the metal by tunneling, or from the semiconductor directly. The arrived electron makes the neighboring cells excited which leads to electron transport to the other cells, as depicted in the figure. The electrons which have traveled throughout the 2D sample, contribute to the in-plane conductivity. The local rules for these electron transmissions are according to the local free energy and the chemical potential.\\
The energy of the electron gas and the chemical potential inside each cell is calculated by means of the Thomas-Fermi-Dirac (TFD) approach. The average energy of the $i$th cell, inside which the charge in supposed to be uniform, is $\left\langle E_i \right\rangle=K(T,\tilde{N}_i)+V_{ee}(T,\tilde{N}_i)+E_{\text{imp}}(T,\tilde{N}_i)$ in which the terms are finite temperature averages of the kinetic, the electron-electron interaction and the impurity energies respectively and $\tilde{N}_i$ is the number of electrons in the cell. In the weak interaction limit (i.e. in which the dominant term is the kinetic energy) $K(T,\tilde{N}_i)$ is readily shown to be $-\frac{2mA\pi}{\beta^2 \hbar^2}\text{Li}_2\left(1-e^{\beta \epsilon_F^i}\right)$ in which $\beta=1/k_BT$, $\epsilon_F^i=\frac{\hbar^2}{2m}\left(\frac{\tilde{N}_i}{\pi A}\right)$, $A$ is the area of the cell which is approximated by $\pi \zeta_{\phi}^2$, $m$ is the electron mass and $\text{Li}_2(z)\equiv\sum_{i=1}^{\infty}\frac{z^n}{2^n}$ is the poly logarithmic function. The inter-particle interaction is composed of two terms: $V_{ee}=\left\langle \hat{V}_{ee} \right\rangle_{\beta}=e^{-\beta E_G}\left\langle \Psi_G\right| \hat{V}_{ee}\left| \Psi_G\right\rangle+\sum_{j\neq G} e^{-\beta E_j} \left\langle \Psi_j\right| \hat{V}_{ee}\left| \Psi_j\right\rangle$ in which $\Psi_G$ is the many body ground state with the energy $E_G$ and $\Psi_j$'s ($j\neq G$) are the many-body excited states with energies $E_j$. For single particle excitations, the contribution of the excitation energies are negligibly small with respect to the ground state contribution so that $V_{ee}(T,\tilde{N}_i)\simeq\left\langle \hat{V} \right\rangle_{T=0}$ ($E_G\equiv 0$). Using the TFD theory and the Gaussian approximation for the two body density matrix, it is calculated to be (for uniform density) $V_{ee}(T,\tilde{N}_i)=\frac{e^2}{8\sqrt{2}\epsilon_0 \zeta_{\phi}}\tilde{N}_i^{\frac{3}{2}}\left( \tilde{N}_i-1\right)^{\frac{1}{2}}\simeq \frac{e^2}{8\sqrt{2}\epsilon_0 \zeta_{\phi}}\tilde{N}_i^2$ for large $\tilde{N}$s \cite{Parr}. The same arguments also hold for the impurity energy whose classical form is $E_{\text{imp}}=-\text{Arcsinh}(1)\frac{Ze^2}{\pi\epsilon_0 \zeta_{\phi}}\tilde{N}_i$ in which $Ze$ is the electric charge of the impurity in the cell and the potential range and its corresponding integrals are supposed to be limited to the cell. The sum of the above mentioned contributions yields the total energy as follows:
\begin{equation}
E_T=\sum_{i=1}^{L}\left[-\alpha T^2\text{Li}_2\left(1-e^{N_i/T}\right) +\beta_0 N_i^2-\gamma_i N_i\right]
\end{equation}
in which $\alpha=2m\left( \frac{\pi k_B \zeta_{\phi}(T)}{\hbar}\right)^2$, $\beta_0=\frac{1}{8\sqrt{2}\epsilon_0 \zeta_{\phi}(T)}\left( \frac{e\alpha}{k_B}\right)^2$, $\gamma_i=\text{sinh}^{-1}(1)\frac{\alpha e^2}{\pi\epsilon_0k_B \zeta_{\phi}(T)}Z_i$, $N_i=\frac{k_B}{\alpha}\tilde{N}_i$ and $L$ is the total number of cells. This is a simplified theory which has the ability to show the non-trivial main features of the experiments, to be described in the following section. \\
The chemical potential of a cell ($\mu_i=\partial A_i/\partial N|_{V,T}$ in which $A_i$ is the Helmholtz free energy of the $i$th cell) as the main building block of the transition rules of electrons between cells is obtained using the relation $A_{\tilde{N}}(V,T)-T\left(\frac{\partial A}{\partial T}\right)_{\tilde{N},V}=\left\langle E\right\rangle$. By considering the fact that $\mu(T\rightarrow 0)\rightarrow 0$ and $ \zeta_{\phi}(T)=aT^{-1/2}$ for two dimensional electron gas \cite{altshuler} ($a$ is a proportionality constant), the solution is obtained to be:
\begin{equation}
\mu_i =k_BT\ln\left(e^{h_i}-1\right)+UT^{\frac{1}{2}}h_i-IZ_iT^{\frac{1}{2}}
\end{equation}
in which $U =\frac{2k_Bm\sqrt{2Da}e^2\pi^2}{8\epsilon_0\hbar^2}$, $I=\text{sinh}^{-1}(1)\frac{e^2}{\pi\epsilon_0\sqrt{Da}}$, $h_i=\frac{N_i}{T}$ and $i$ stands for the $i$th cell. By setting $U=0$ we show that the existence of the localized-extended phase transition in our system is not interaction-driven. Instead the effect of randomness of $Z_i$'s, which captures the on-site (diagonal) disorder is investigated. $Z_i$'s are supposed to be random noise with an uniform probability measure $P(Z)=\frac{1}{\Delta}\Theta(\Delta/2+(Z-Z_0))\Theta(\Delta/2-(Z-Z_0))$ in which $\Delta$ shows the disorder strength, $Z_0=\left\langle Z\right\rangle$ is the average of $Z$ and $\Theta$ is the step function and $\left\langle Z_iZ_j\right\rangle=\delta_{ij}$ in which $\left\langle \right\rangle $ shows the ensemble average and $\delta$ is the Kronecker delta. \\

The construction of the local transition rules needs the information of the charge content and the local chemical potential of the donor (original) cell (say cell $1$ with local chemical potential $\mu_1$) and the acceptor (destination) cell (say cell $2$ with local chemical potential $\mu_2$), both embedded in a large system with the chemical potential $\mu_0$. The relative probability of this transition ($P_{1\rightarrow 2}$) is then proportional to $e^{-\beta(\mu_1-\mu_2)}$ and the probability measure of adding an electron to the system at the $i$th cell is proportional to $e^{-\beta(\mu_0-\mu_i)}$. Therefore the charge density configuration $\left\lbrace \tilde{N}_i\right\rbrace_{i=1}^L$ determines totally the behavior of the system in the next time. Accordingly we introduce a cellular automaton model with simple local rules for the charge transfers between the cells. Consider a cell ($0$) with its neighbors each of which has its own local chemical potential $\mu_i$, $i=1,2,...,6$ in such a way that $\mu_i<\mu_j$ for $i<j$. The site $0$ is said to be \textit{unstable} (has the potential to give an electron to its neighbors) if its chemical potential $\mu(0)$ exceeds the chemical potential of the bulk $\mu_0$. In this case one checks the probability of the electron transfer to each of its neighbors. This probability is firstly checked for a neighboring site with the lowest chemical potential, namely $i=1$. Afterwards the next candidate for the electron transfer is the site with the nearest $\mu$ to $\mu_1$, i.e. $\mu_2$ etc. The electron transmission from $0$ to the neighboring site $i$ is occurred with the probability $P_{0\rightarrow i}\sim \Theta(\mu(0)-\mu_0) \times\text{Max}\left\lbrace 1,e^{-\beta\left( \mu(i)-\mu(0)\right)}\right\rbrace$. The electrons also can leave the system via the boundaries by assigning the chemical potential $\mu_0$ to the (virtual) external cells which can be interpreted as the work function of the system. \\

\begin{figure*}
	\centerline{\includegraphics[scale=4.5]{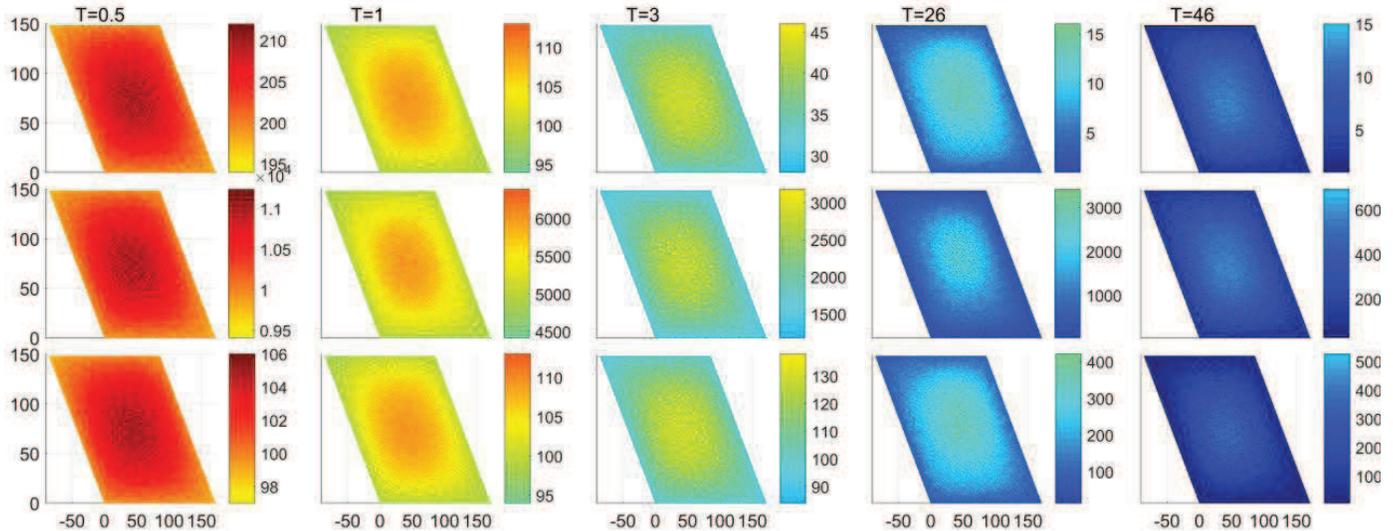}}
	\caption{Color online. The contour plot of the charge density (first row), the energy (second row) and chemical potential (third row) profile of a disorder-free ($\Delta=0$) system with various temperatures. It is seen that the charge density decreases with temperature.}
	\label{fig:chargecontourplot}
\end{figure*}

\begin{figure*}
	\begin{subfigure}{0.45\textwidth}\includegraphics[width=\textwidth]{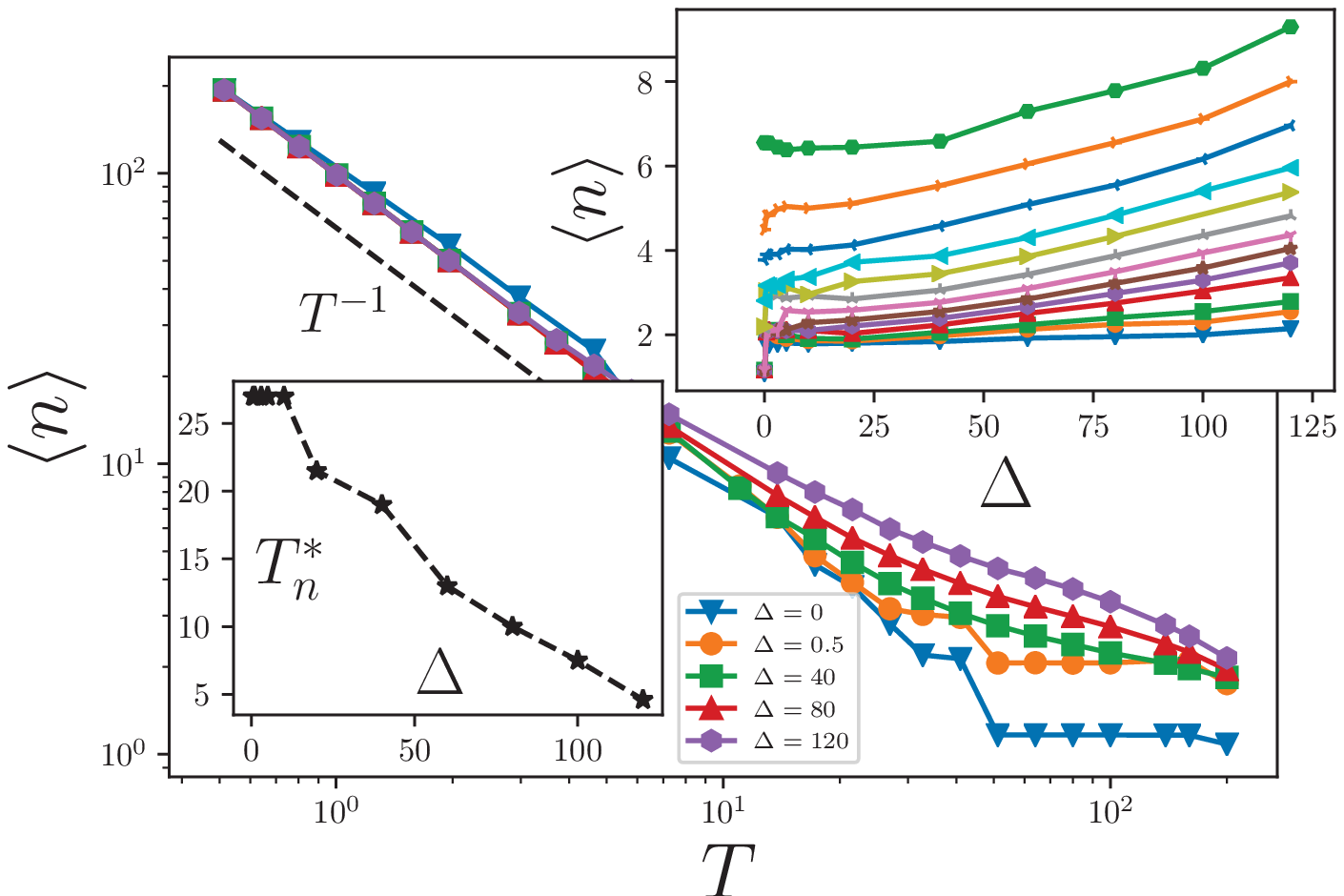}
		\caption{}
		\label{fig:DensityInTermTem}
	\end{subfigure}
	\begin{subfigure}{0.45\textwidth}\includegraphics[width=\textwidth]{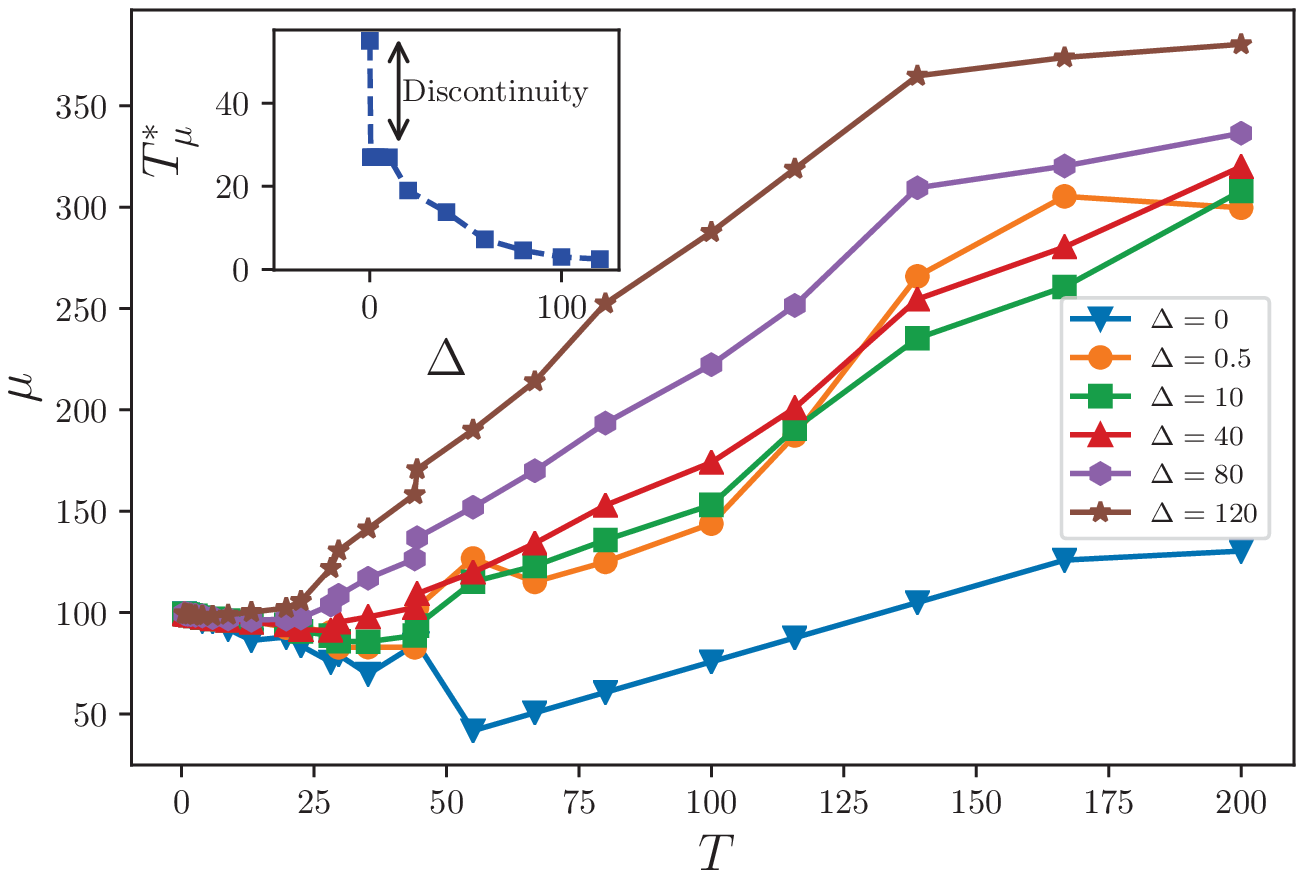}
		\caption{}
		\label{fig:muInTermTemp}
	\end{subfigure}
	\begin{subfigure}{0.45\textwidth}\includegraphics[width=\textwidth]{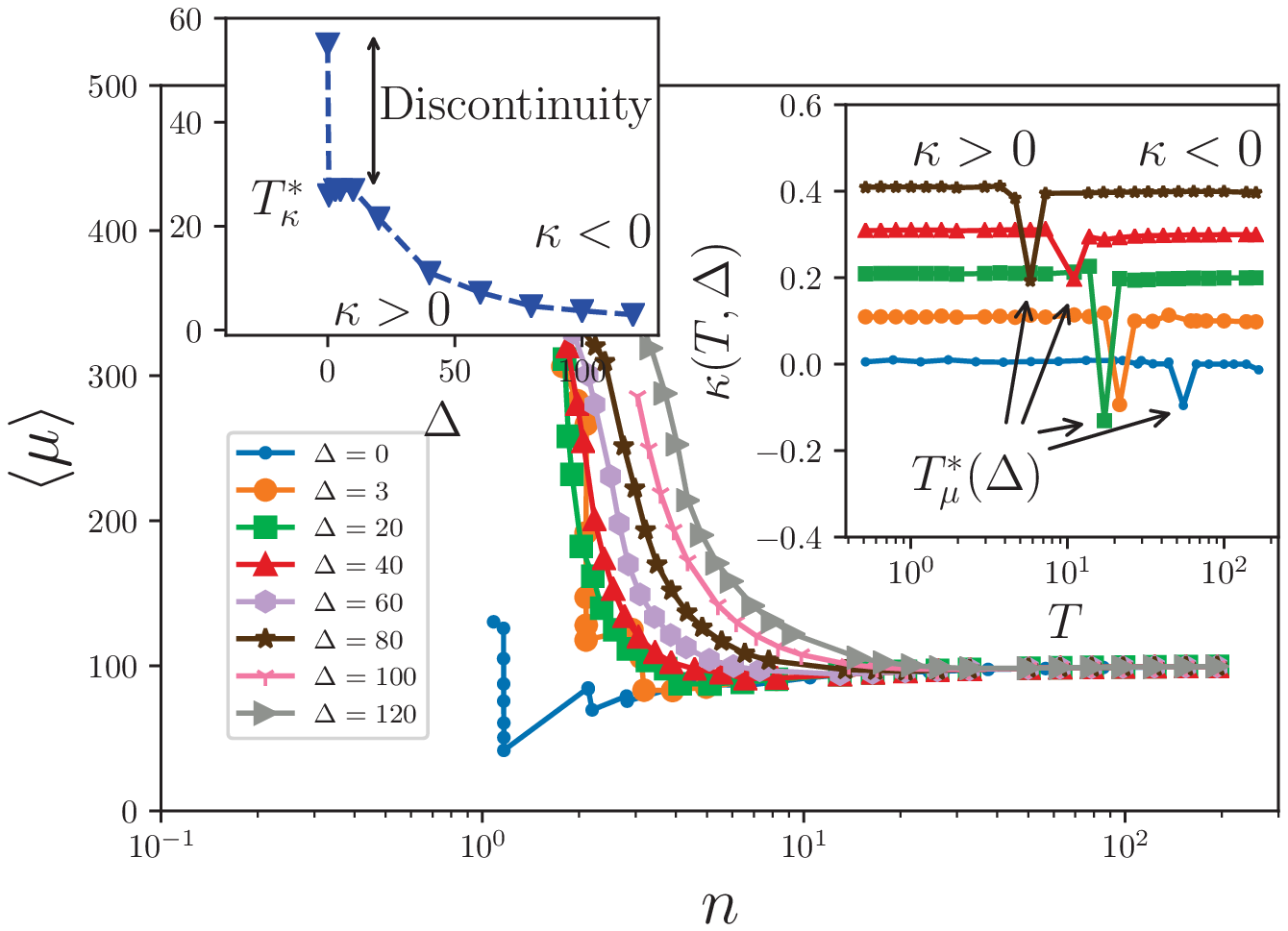}
		\caption{}
		\label{fig:mu-n}
	\end{subfigure}
	\begin{subfigure}{0.45\textwidth}\includegraphics[width=\textwidth]{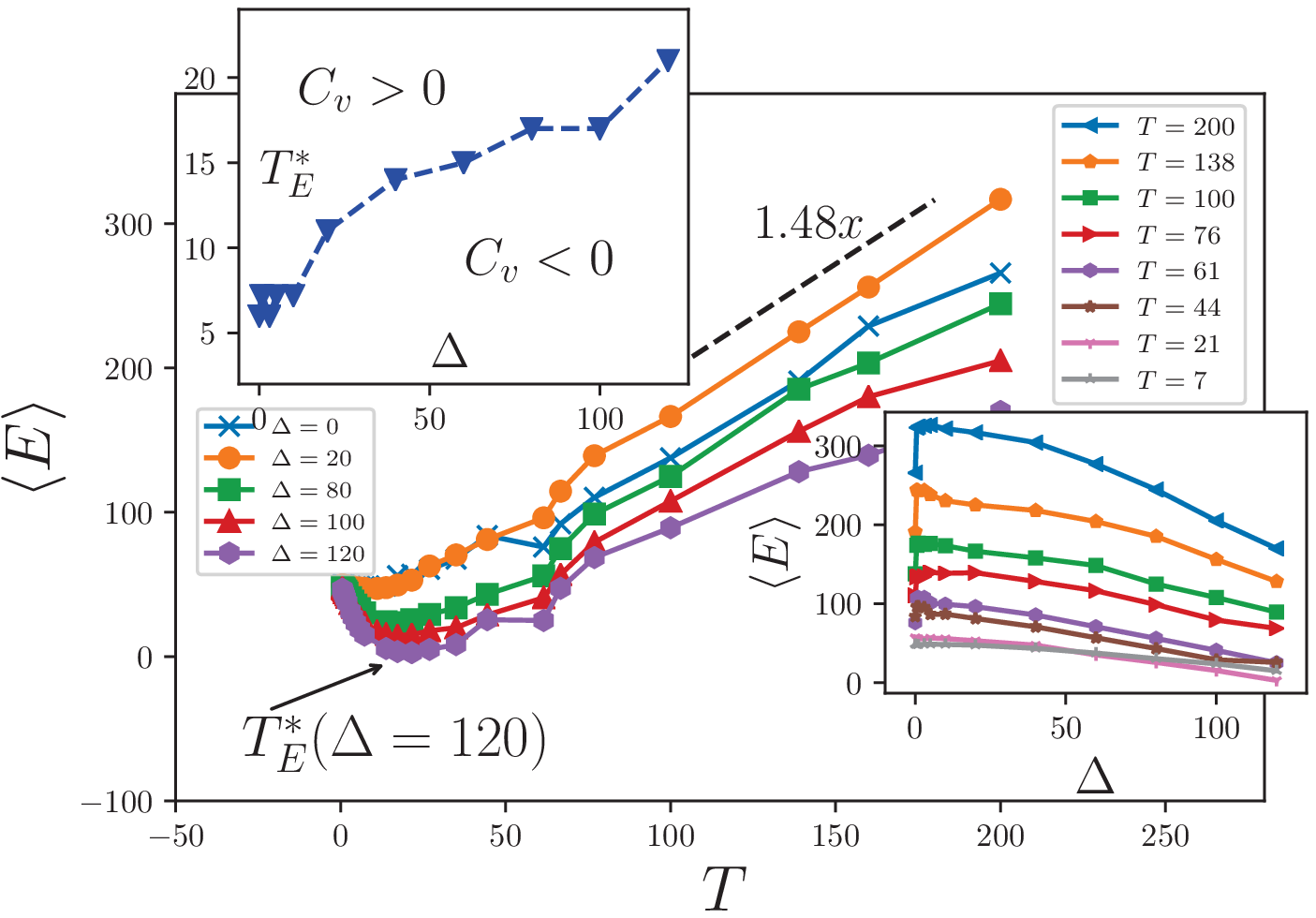}
		\caption{}
		\label{fig:EnergyInTermTemp}
	\end{subfigure}
	\caption{Color online (a) Power-law behavior of the average density in terms of the temperature for $\Delta=0-120$ (upper inset: $\left\langle n\right\rangle $ in terms of the $\Delta$ for $T=11-200$ from top to bottom, lower inset: $T^*_n$ in terms of $\Delta$). (b) The dependence of the chemical potential on $T$, inset: $T^*_{\mu}$ in terms of $\Delta$. (c) The dependence of the chemical potential on $\left\langle n\right\rangle$ for $\Delta=0-120$, left inset: $T^*_{\kappa}$ in terms of $\Delta$, right inset: shifted $\kappa$ in terms of $T$ for $\Delta=0-80$. $T^*_{\kappa}$ is defined as the temperature in which the compressibility diverges. (d) The dependence of the mean energy on $T$ for $\Delta=0-120$, upper inset: $T^*_E$ in terms of the $\Delta$, lower inset: $\left\langle E\right\rangle$ in terms of $\Delta$ for $T=7-200$ from bottom to top.}
	\label{LocalObservables}
\end{figure*} 

The dynamics is started by injecting an electron at a random cell $i_0$ of a randomly distributed electronic system from the charge reservoir. If the mentioned cell becomes unstable ($\mu(i_0)>\mu_0$) the electron transfer to the neighbors is checked and occurs if permissible, which may cause the destination cells to become unstable. Therefore a chain of charge transfers may occur through the system. This continues until all cells over the system become stable. Then another random cell is chosen for the injection. Although after a chain of local relaxations there is no unstable cell, the system may be far from the equilibrium since the charge transfer is a non-equilibrium process. Therefore the system should find the configuration with lowest free energy at each step (equilibration). For this purpose some random cells are chosen for testing whether the charge transfer to its neighbors is favorable or not. In our analysis, after $n=\frac{1}{10}L_x\times L_y$ ($n=$ number of electron injections) \textit{fast relaxations}, we do $100L_x\times L_y$ searches for favorable particle transfers (local equilibration). The overall process may be divided into two distinct behaviors: first the stage at which $\left\langle \mu\right\rangle <\mu_0$ (SI state), and secondly the stage at which $\left\langle \mu\right\rangle \simeq\mu_0$ (SII state). In SI $\left\langle \mu\right\rangle$ grows linearly with $n$ up to a step $n_0$ at which $\left\langle \mu\right\rangle \simeq\mu_0$. In this stage the system is non-equilibrium. In SII however, $\left\langle \mu\right\rangle$ is expected to become statistically constant, the number of injections is statistically equal to the number of electrons which leave the system. All of our analysis is in the SII stage.

\section{Singular behaviors and local phase transitions}\label{local}

The properties of the electronic system are investigated in terms of $T$ and $\Delta$, and the above mentioned second order percolation transition is characterized in this section. All $T$ and $\Delta$ values in this part are in arbitrary units, thus the sole values are not important, we instead emphasis on the trends and behaviors. The contour plots in the Fig. \ref{fig:chargecontourplot} help to visualize the charge density, the energy and the chemical potential profiles of the electrons ($n(\textbf{r})$, $E(\textbf{r})$ and $\mu(\textbf{r})$ respectively), in which it is evident that the average density $\left\langle n\right\rangle $ decreases with the temperature. The fact that $\left\langle n\right\rangle $ monotonically decreases with temperature in a power-law fashion ($T^{-1}$) is evident in Fig. \ref{fig:DensityInTermTem} for $\Delta=0$, showing that the frequency of leaving of the electrons via the boundaries is larger for the higher temperatures. For $\Delta>0$ there is a temperature at which the corresponding graph deviates from the (linear part of the) $\Delta=0$ graph which is named as $T^*_n(\Delta)$. $T^*_n$ has non-monotonic dependence on $\Delta$ (lower inset of Fig. \ref{fig:DensityInTermTem}). It is constant up to $\Delta\simeq 10$ and decreases monotonically afterwards, showing that the disorder facilitates the transition to the new regime. The fact that $\left\langle n\right\rangle$ is an increasing function of $\Delta$ (the upper inset) shows that the disorder has a trapping role for the electrons, preventing them to leave the system from the boundaries. To search more deeply for the physics of this change of behavior, one may calculate the statistics of the chemical potential $\mu$ and the compressibility ($\kappa$) which have been shown in Figs. \ref{fig:muInTermTemp} and \ref{fig:mu-n}. $T^*_n$ is accompanied with two other temperature scales ($T^*_{\mu}$ and $T^*_{\kappa}$) in which some singularities arise. $T^*_{\mu}$ is defined as the point in which a negative-positive slope transition of the average chemical potential ($\text{d}\left\langle \mu\right\rangle/\text{d}T>0 \rightarrow \text{d}\left\langle \mu\right\rangle/\text{d}T<0$) occurs. The behavior of $\left\langle \mu \right\rangle $ in terms of $\left\langle n\right\rangle $, i.e. Fig. \ref{fig:mu-n} yields some other important data, e.g. it becomes constant for high densities, and its slope vanishes at some $\Delta$-dependent temperature, namely $T^*_{\kappa}$. At this temperature the compressibility which is defined by $\kappa^{-1}=N^2\left( \frac{\partial \mu}{\partial N}\right)_{V,T}$ becomes divergent. It is interestingly seen that $T^*_n$, $T^*_{\mu}$ and $T^*_{\kappa}$ have more or less the same dependence on $\Delta$, showing that we are observing a unique physical phenomenon, which is shown (in the next section) to be a geometric percolation transition. A different behavior was obtained for $\left\langle E\right\rangle$ which is nearly linear in $T$ with positive slope for high temperatures, but becomes decreasing function of the temperature for low temperatures, i.e. below a critical temperature shown by $T^*_E$. This shows that the electron gas becomes unstable. $T^*_E(\Delta)$ is an increasing function of $\Delta$ which opposes the other critical temperatures. We interpret $T^*_E$ as the extent below which our TFD treatment is not valid.\\
The question whether the disorder can be processed perturbatively or not, can be addressed by investigating the properties of the model in going from $\Delta=0$ to positive (small) $\Delta$ values. It is worthy to note that the behavior of $\Delta=0$ and low temperatures is very different from other $\Delta$ values. The isolated behavior of the regular system ($\Delta=0$) (Figs. \ref{fig:DensityInTermTem}, \ref{fig:muInTermTemp} and \ref{fig:mu-n}) and the observation that an abrupt jump occurs for all critical temperatures $T^*_n$, $T^*_{\mu}$ and $T^*_{\kappa}$ from $\Delta=0$ to non-zero $\Delta$'s (i.e. a discontinuity) show that the disorder is a non-perturbative quantity in our model, i.e. one cannot perturbatively approach $\Delta=0^+$ from $\Delta=0$. 

\section{Global properties and the percolation transition}\label{global}

\begin{figure*}
	\begin{subfigure}{0.45\textwidth}\includegraphics[width=\textwidth]{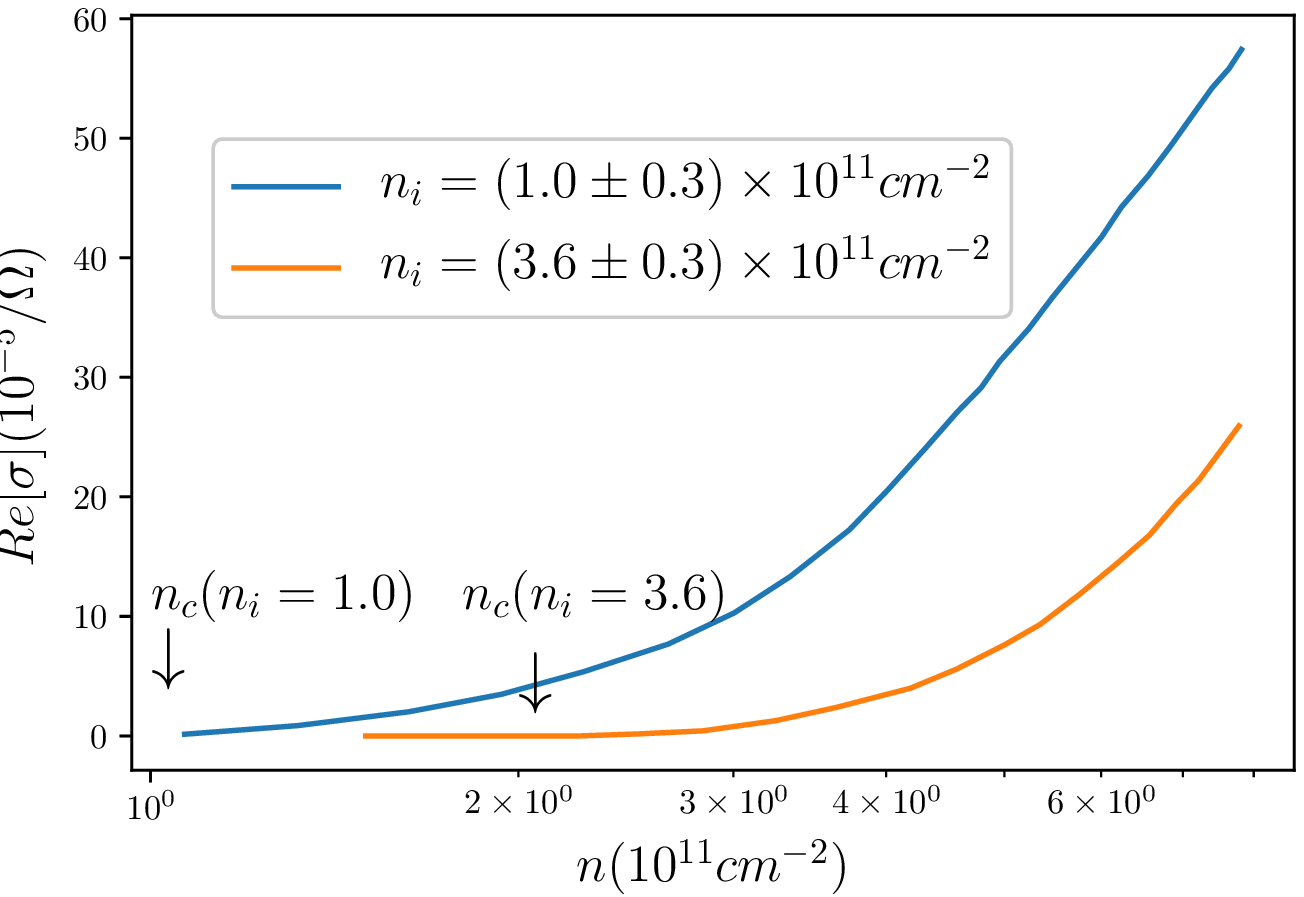}
		\caption{}
		\label{fig:ConductivityExperiment}
	\end{subfigure}
	\begin{subfigure}{0.5\textwidth}\includegraphics[width=\textwidth]{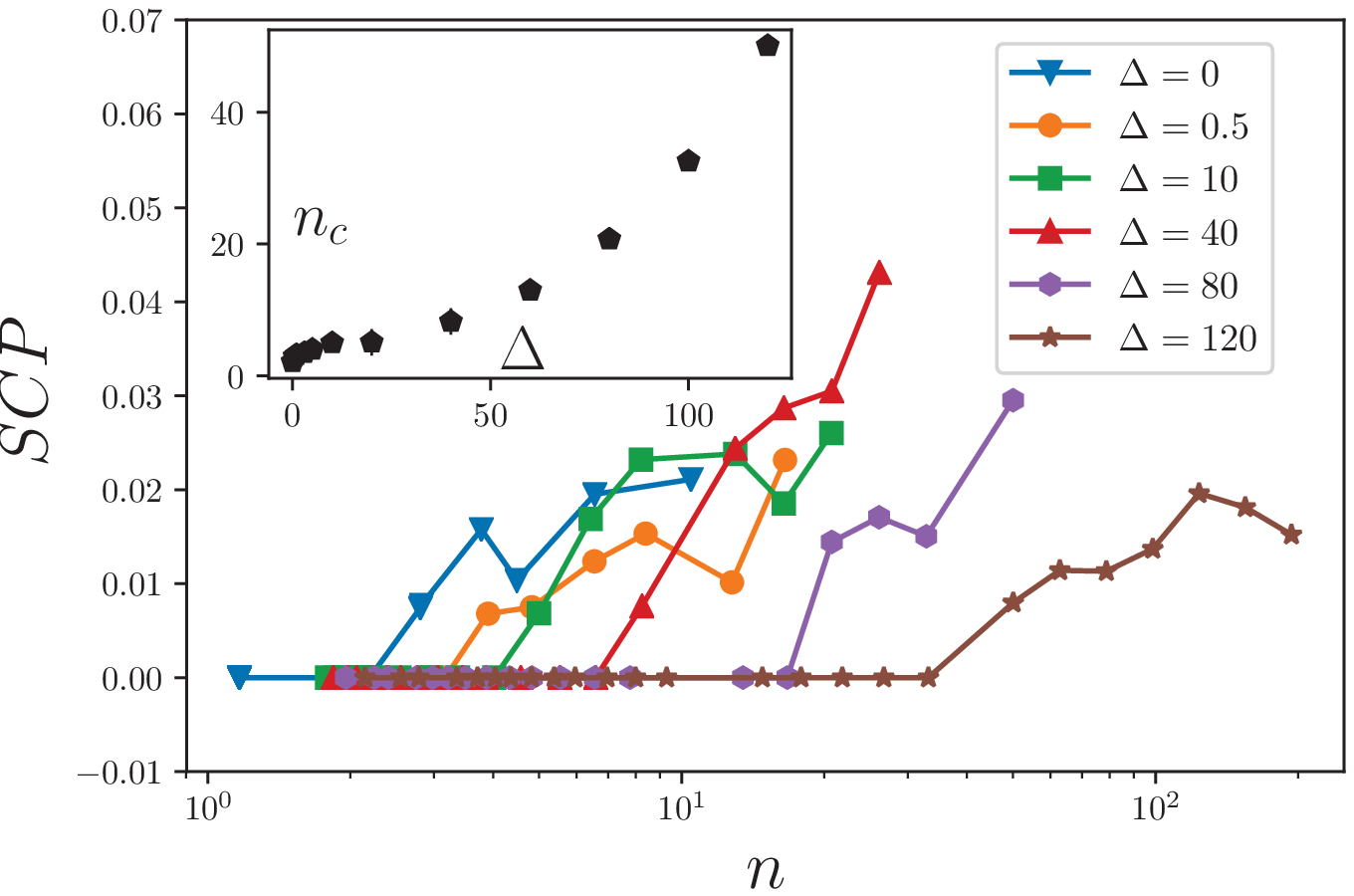}
		\caption{}
		\label{fig:Perc-Density}
	\end{subfigure}
	\begin{subfigure}{0.48\textwidth}\includegraphics[width=\textwidth]{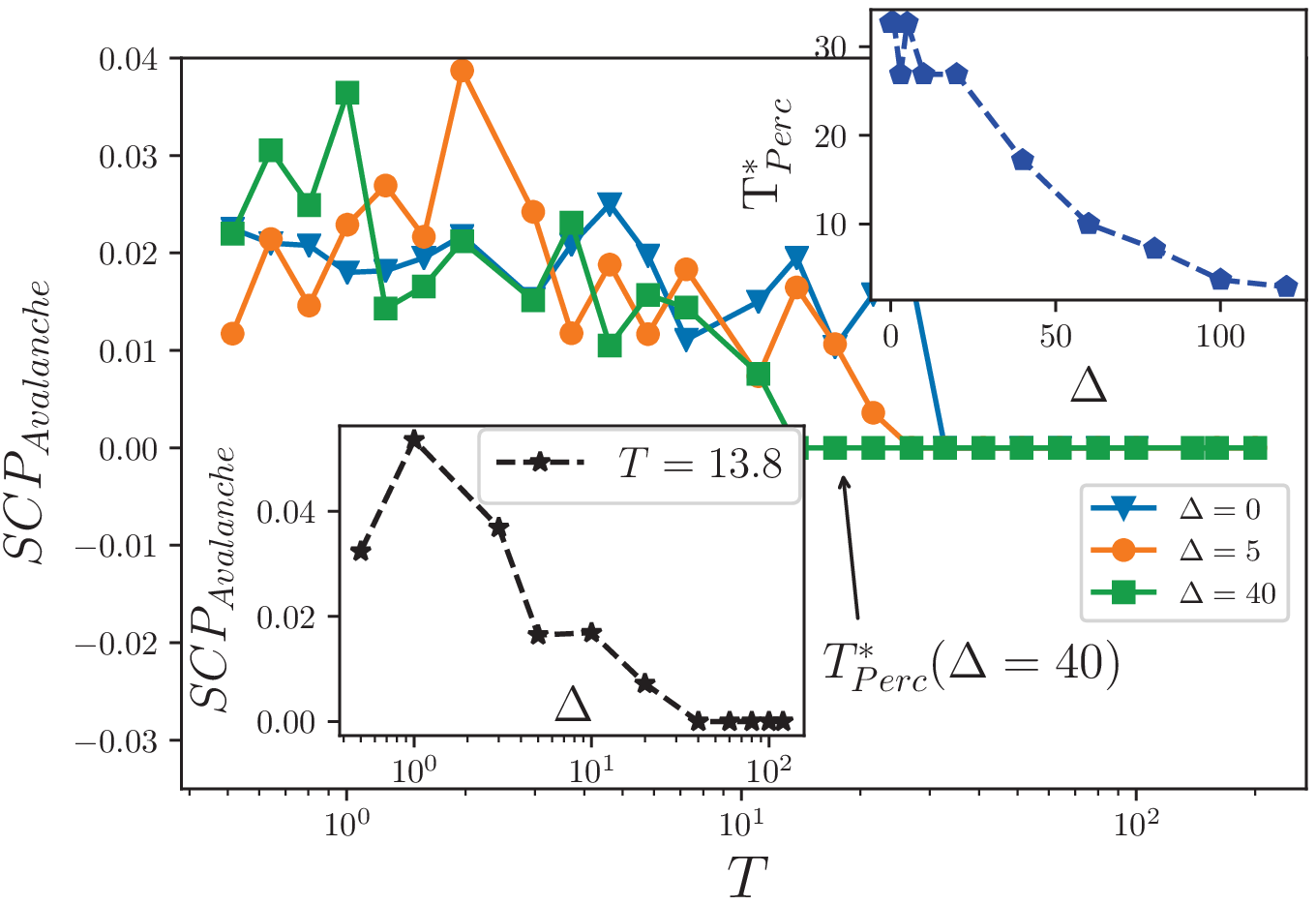}
		\caption{}
		\label{fig:PercInTermTemp}
	\end{subfigure}
	\begin{subfigure}{0.48\textwidth}\includegraphics[width=\textwidth]{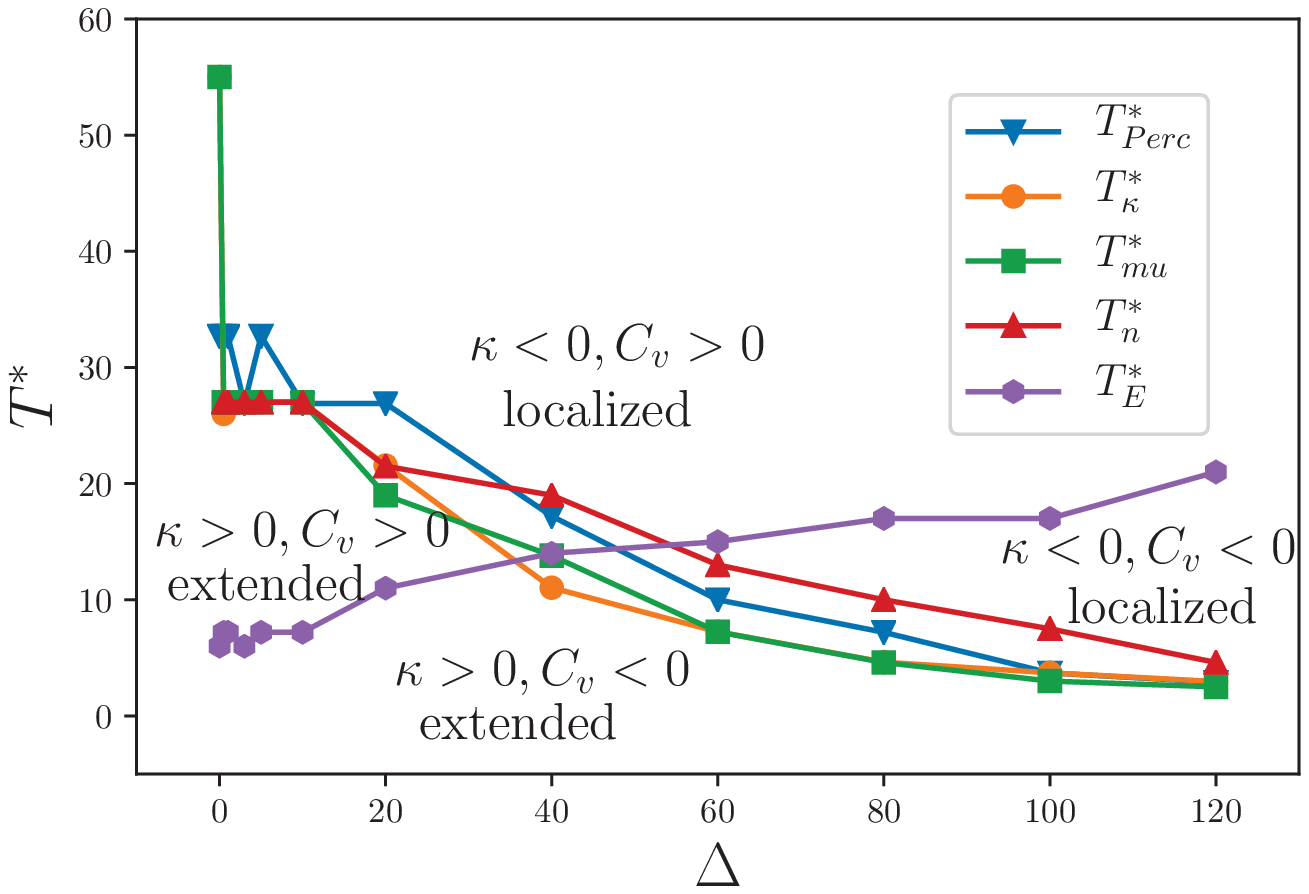}
		\caption{}
		\label{fig:TotalCriticalTemp}
	\end{subfigure}
	\caption{color onloine (a) The experimental and theoretical results for (real part of) the static conductivity as a function of electron density for two impurity densities $n_i=1.4\times 10^{11}\text{cm}^{-2}$ and $n_i=3.4\times 10^{11}\text{cm}^{-2}$: Gold, Gotze, Mazure, and Koch \cite{GoldLocalizationExp} measured at $\omega=0.001\text{cm}^{-1}$. (b) The SCP in terms of the density $n$ for various rates of $\Delta$, inset: the critical density in terms of $\Delta$. (c) The decreasing behavior of the SCP in terms of the temperature (metallic behavior) for various rates of $\Delta$, lower inset: SCP in terms of $\Delta$  for $T=13.8$ which vanishes for some $\Delta$. Upper inset: $T^*_{perc}$ in terms of $\Delta$. (d) The total phase diagram with two separate phases in which  it is seen that all $T^*_n$, $T^*_{\mu}$, $T^*_{\kappa}$ and $T^*_{perc}$ are more or less the same.}
	\label{SCPs}
\end{figure*}

The investigation of the global quantities yields deeper insight to the physics of the transition line. The power-law behaviors of the (geometrical and local) observables show that the system lacks any preferred length scale due to diverging the correlation length of the system. In the geometrical approach, we deal mostly with the \textit{electronic avalanches} which is defined as the chain of relaxations between two successive stable configurations. Here we consider the set of cells which have had at least one toppling ($\equiv$ a relaxation process) in an avalanche. A \textit{spanning avalanche} or \textit{cluster} is an avalanche area that connects two opposite boundaries. When a spanning cluster (avalanche) occurs, the moving electrons which are involved in the avalanche process have passed throughout the sample which results to the electronic conduction. We define the spanning cluster probability (SCP) which is the probability that a random chosen cell belongs to a spanning cluster. A laboratory observable that is expected to directly be related to the SCP is the system conductivity, since it is related to the ability of transferring the electrons from one side to the opposite side of the system. The real part of the conductivity (at fixed frequency) of 2D Si(100) metal-oxide-semiconductor disordered by Coulomb impurities (Na ions) in zero temperature was shown to vanish at some (charged) impurity-dependent critical density $n_c$ as is depicted in Fig. \ref{fig:ConductivityExperiment} (by Gold \textit{et. al.}~\cite{GoldLocalizationExp}) and theoretically confirmed \cite{GoldLocalization}. This transition is disorder-driven, since the disorder pushes the transition point to the larger values, and is second order, since the relation is power-law in the vicinity of the transition point. Our model (Fig. \ref{fig:Perc-Density}) yields the main features of this experimental observation. The SCP shows a transition from $n\leq n_c$ (in which $SCP=0$) to the phase $n>n_c$ in which $SCP>0$. The disorder also pushes $n_c$ to the larger values in the same fashion of the experimental observations, which has been shown in the inset. The fact that in the $n>n_c$, the system shows \textit{metallic behaviors}, can be deduced from the Fig. \ref{fig:PercInTermTemp} in which for $T<T^*_{\text{Perc}}$ (corresponding to $n>n_c$) $\frac{\text{d}}{\text{d}T}SCP<0$. SCP has also a decreasing behavior in terms of $\Delta$ and vanishes at some $\Delta$ for fixed temperature, showing that the transition is disorder-driven. Therefore our simple percolation model yields the main features of the experimental results, highlighting the finite temperature percolation model as a candidate. More interestingly we see from the upper inset of this figure that $T^*_{\text{Perc}}$ has the same behaviors in terms of $\Delta$, as $T^*_n$, $T^*_{\mu}$ and $T^*_{\kappa}$, Fig. \ref{fig:TotalCriticalTemp} in which the total phase diagram of the system has been shown. This fitness, shows that they are the same physical phenomenon, i.e. the percolation transition between extended-non-extended phases.\\
\begin{figure*}
	\begin{subfigure}{0.43\textwidth}\includegraphics[width=\textwidth]{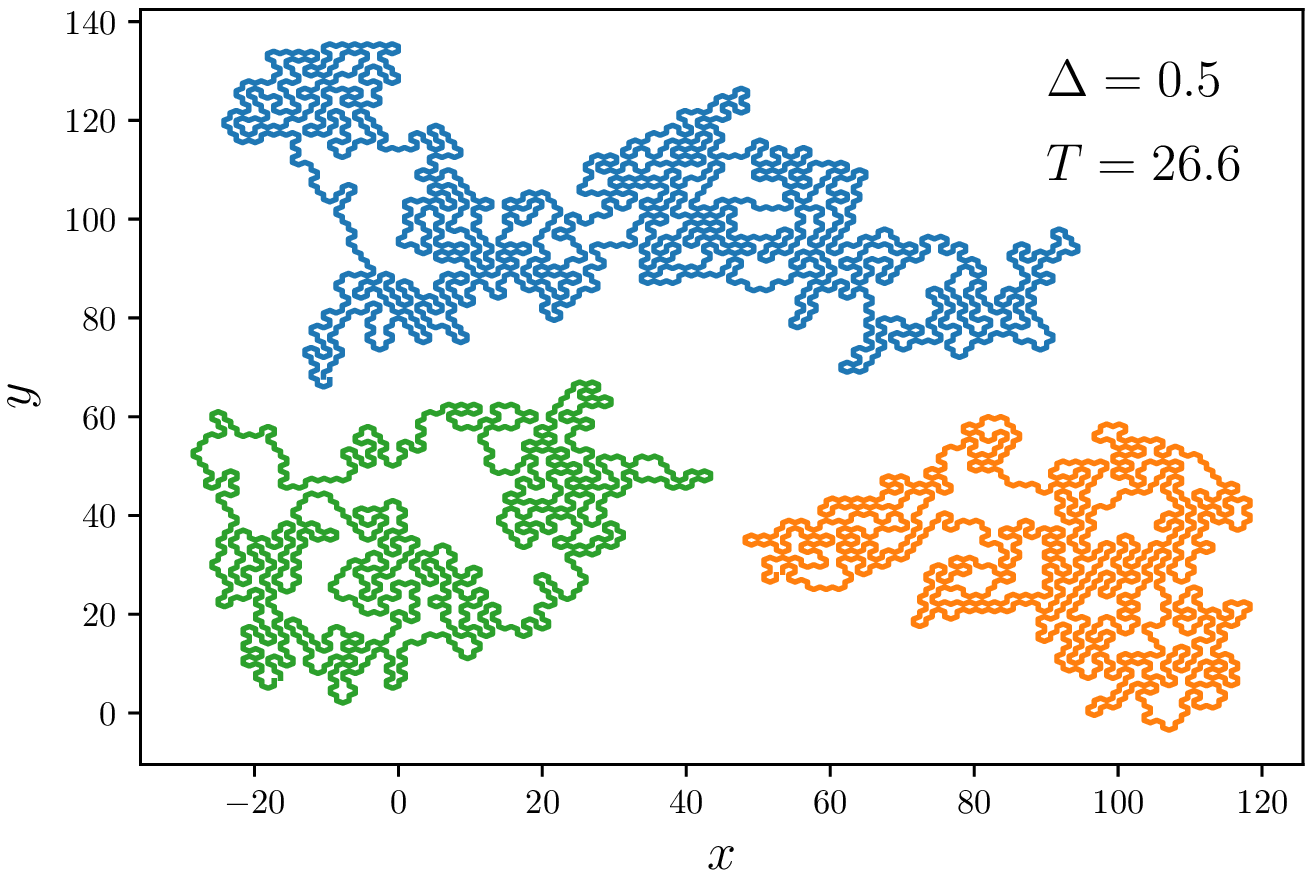}
		\caption{}
		\label{fig:Domainwall-Slope1}
	\end{subfigure}
	\begin{subfigure}{0.48\textwidth}\includegraphics[width=\textwidth]{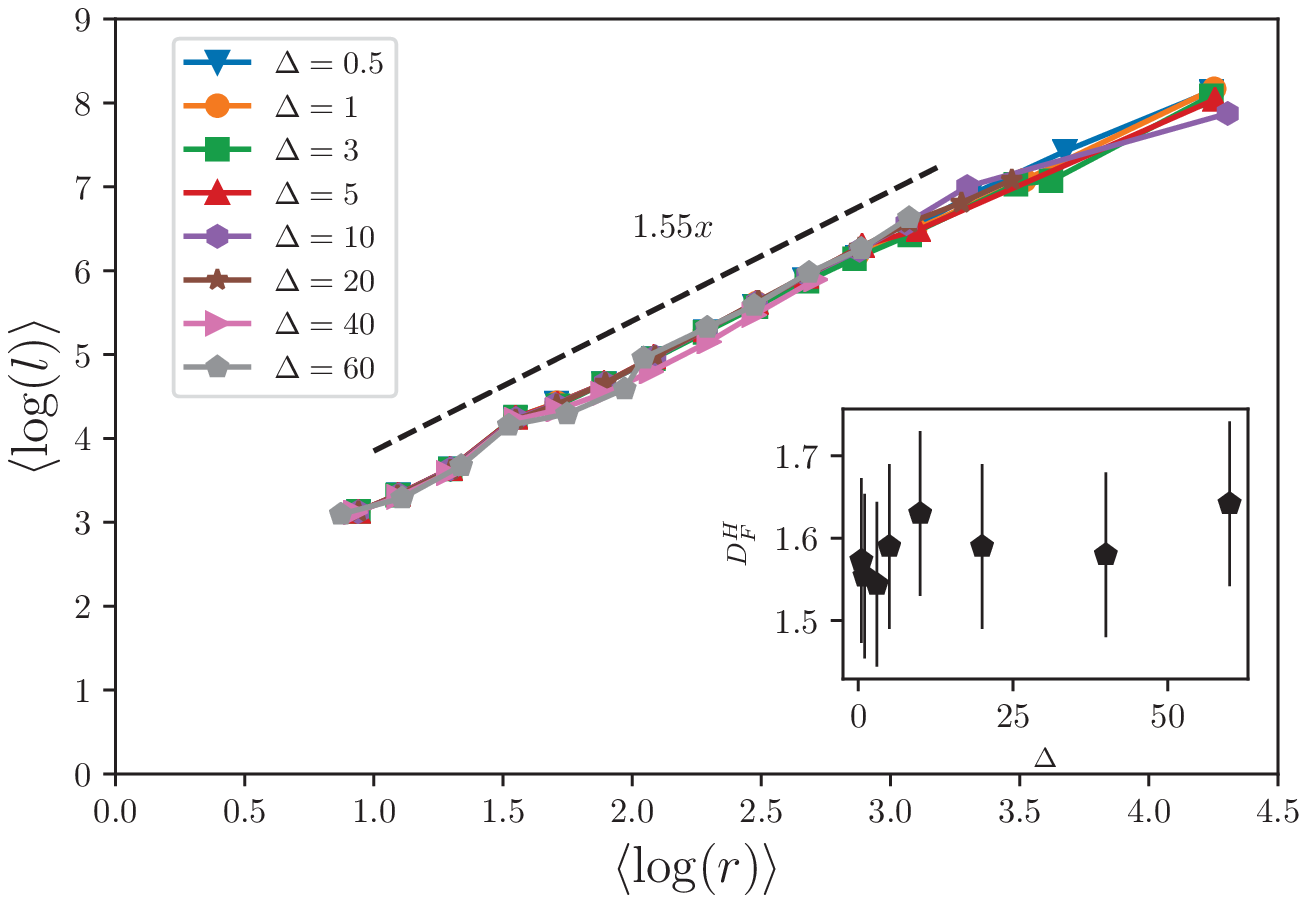}
		\caption{}
		\label{fig:FDSlope}
	\end{subfigure}
	\begin{subfigure}{0.45\textwidth}\includegraphics[width=\textwidth]{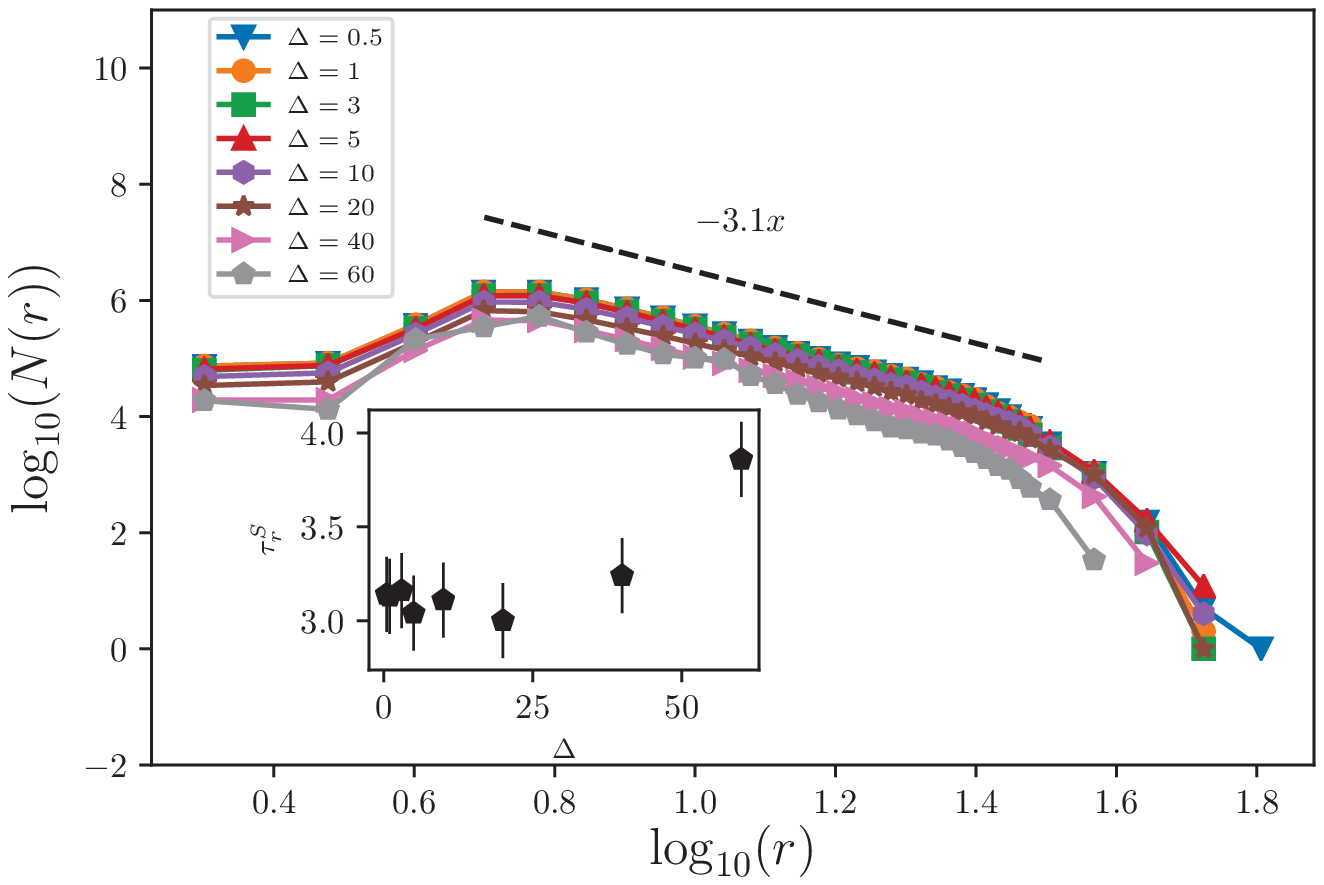}
		\caption{}
		\label{fig:GyrSlope}
	\end{subfigure}
	\begin{subfigure}{0.45\textwidth}\includegraphics[width=\textwidth]{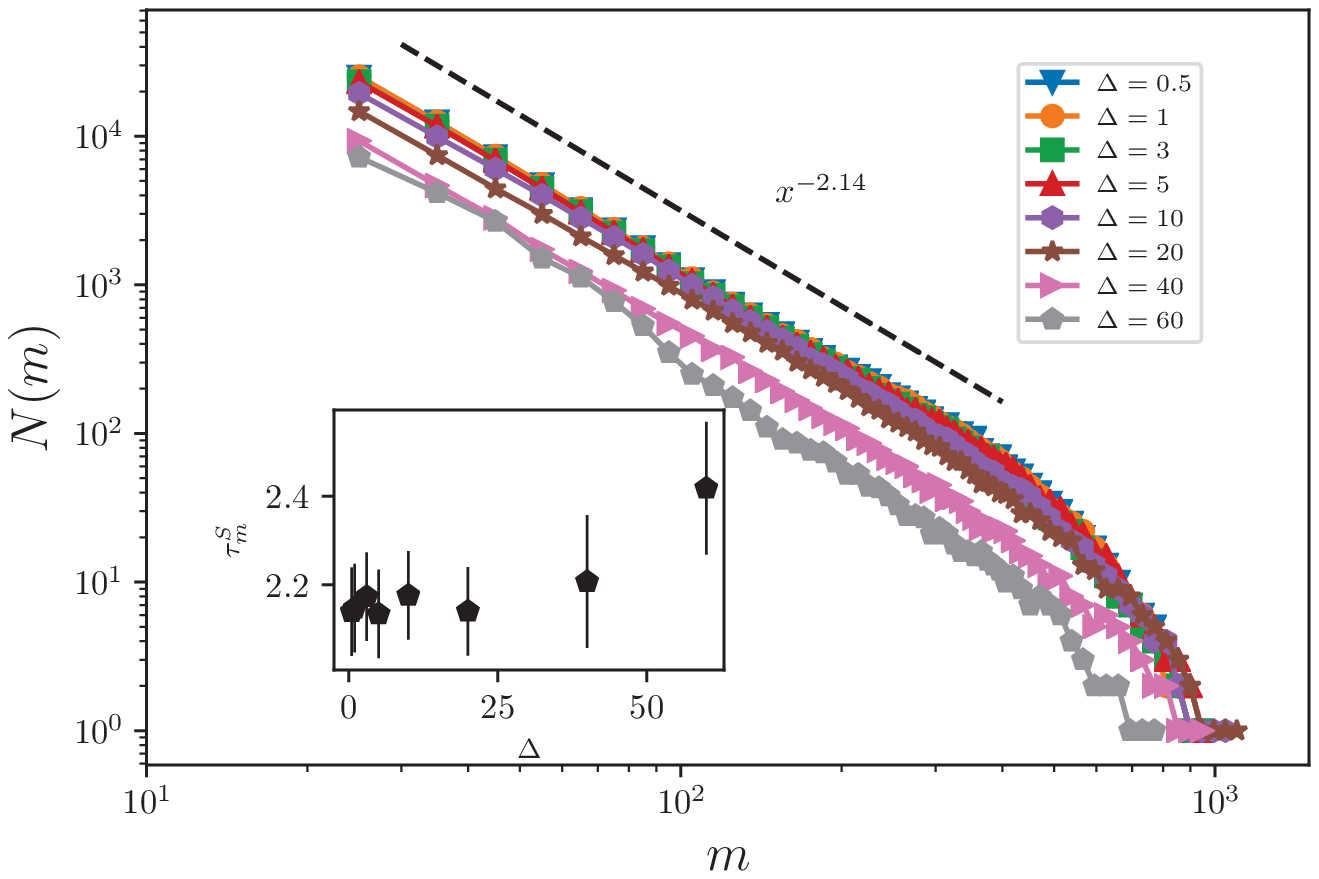}
		\caption{}
		\label{fig:LoopsSlopeMass}
	\end{subfigure}
	\caption{Color online. (a) the external frontiers of a $S-$field sample for $\Delta=0.5$ and $T=26.6$. (b) The fractal dimension of the $n-$field loops. Inset: The fractal dimension in terms of $\Delta$. (c) The distribution function of the gyration radius of the loops of the $S-$field. Inset: the exponent $\tau_r^S$ in terms of $\Delta$. (d) The distribution function of the mass of the avalanches ($m$) of the $S-$field. Inset: The exponent $\tau_m^S$ in terms of $\Delta$.}
	\label{GlobalObservables}
\end{figure*}

The power-law behavior of the system cannot be concluded form the calculated SCP, due to its large error bars in the extended phase. This can be tested however in the other observables which characterize the transition. Each avalanche has its own \textit{toppled area} ($m$), the external frontier which forms a loop (with length $l$) and the gyration radius ($r$). Consider the charge density field $n(x,y)$ (named as $n$-field) and its contours. We define the slope field $S(x,y)$ (named as $S$-field) by the directional derivative along the unit vector $\textbf{e}_r$, $S(x,y)\equiv\nabla n.\textbf{e}_r=\frac{1}{r}\left(x\partial_x+y\partial_y\right)n(x,y)$, in which $\textbf{n}$ is the unit vector along the line extended from the origin (which is supposed to be $(L_x/2,L_y/2)$) to the point $(x,y)$. To extract these loops we have used Hoshen-Kopelman algorithm \cite{Hoshen}. A $S$-field sample has been shown in Fig. \ref{fig:Domainwall-Slope1}. In the critical state the distribution function ($N$) of all quantities are expected to show power-law behaviors and the observables scale with each other by some exponents which are the fractal dimensions. The fractal dimension of loops $D_F$ for the $n$-field that is defined by $\left\langle \ln(l)\right\rangle=D_F\left\langle \ln(r)\right\rangle$ has been shown in Fig. \ref{fig:FDSlope}. The $n$-filed exponents are $\Delta$-dependent, whereas the $S$-field exponents are universal ($\Delta$-independent) up to $\Delta^*$ in which the $\kappa=0$ line meets the $c_V=0$ line (see Figs.\ref{fig:GyrSlope} and \ref{fig:LoopsSlopeMass} and the TABLE \ref{tab1}). Interestingly this is compatible with the Kravchenko's observation that \textit{for the small values of $\Delta$ the critical behaviors are universal, whereas for the highly disordered samples are not}~\cite{Kravchenko}.\\
\begin{table}[h]
	\begin{tabular}{ l | c | c | c | c }
		\hline 
		& $D_F$ & $\tau_l$ & $\tau_m$ & $\tau_r$ \\
		\hline
		$S-$filed & $1.7\pm 0.02$ & $2.3\pm 0.1$ & $2.1\pm 0.1$ & $3.1\pm 0.3$ \\
		$n-$field & $1.55\pm 0.02$ & $\Delta$-dependent & $\Delta$-dependent & $\Delta$-dependent \\
		\hline 
	\end{tabular}
	\caption{The geometrical exponents of the MIT transition.}
	\label{tab1}
\end{table}
The fractal dimension of the $n$-field is compatible with Gaussian free field (GFF) $D_F^{\text{GFF}}=\frac{3}{2}$, whereas the fractal dimensional of the $S$-field is mostly compatible with the ordinary critical percolation theory $D_F^{\text{percolation}}=\frac{7}{4}$.

\section{Discussion}

The paper has been devoted to developing a quantum automaton model for the 2D transport of the electrons in a 2D system in contact with an electronic reservoir. It was argued that the system can be meshed by some cells of the linear size of the phase-relaxation length which is temperature dependent. The temperature and density-dependent free energy of the cells was obtained using the Thomas-Fermi-Dirac (TFD) approach. The transition of electrons between cells is permissible by measuring the relative probability which is related to the local chemical potential of each cell. We obtained the properties of the system in terms of temperature $T$ and disorder strength $\Delta$ in the zero inter-particle interaction limit. It was shown that the disorder is non-perturbative , i.e. $\Delta= 0^+$ can not be obtained perturbatively from $\Delta=0$. By investigating the local quantities of the system we found three energy scales, namely $kT^*_n$, $kT^*_{\mu}$ and $kT^*_{\kappa}$ which were shown to be (more or less) the same. At this transition line the compressibility vanishes.\\
The global investigation revealed that there is another energy scale $kT^*_{\text{Perc}}$ in which a percolation transition between extended-non-extended phases occur. The disorder pushes the critical density $n_c$ to the larger values, i.e. $n_c(\Delta)$ is an increasing function of $\Delta$ in agreement with the experimental observations. It was seen that this percolation transition line coincides with the zero-compressibility line, revealing that all of these observations show a single physical phenomenon. In the extended phase the spanning cluster probability (which is directly related to the conductivity) decreases with increasing temperature which is a metallic behavior. \\
The order of the transition was argued to be two, for which some power-law behaviors arise. The geometrical statistical observables were calculated for two kind of fields, namely the $n$-field (the charge density field) and the $S$-field (the slope of the $n$-field), each of which has its own critical exponents in the percolation transition line. The exponents of the $n$-filed were shown to be disorder- dependent, whereas the critical exponents of the $S$-field are disorder-independent up to some $\Delta^*$ in accordance with the Kravchenko's observation~\cite{Kravchenko,Kravchenko1} in which it is stated that the exponents are universal for low disorder values, and are non-universal for large disorders. The fractal dimension of the loops of $n$-filed are mostly compatible with the Gaussian free field (GFF) theory, i.e. $D_F^{\text{GFF}}=\frac{3}{2}$. The exponents of the $S$-field however accord with the percolation theory $D_F^{2D\text{percolation}}=\frac{7}{4}$.

\end{document}